\documentclass[aps,prl,twocolumn,groupedaddress]{revtex4}
\usepackage{amssymb,amsfonts,amsmath,graphicx,color,mathptm,times}
\usepackage{mathrsfs,natbib}
\usepackage{amsmath,amssymb, amsfonts,bbm}
\usepackage{graphics,epsfig,color,subfigure}
\usepackage{hyperref}
\usepackage{verbatim}
\usepackage{eufrak}
\usepackage{natbib}

\newcommand{\be}{\begin{equation}}
\newcommand{\ee}{\end{equation}}

\def\bea{\begin{align}}
\def\ena{\end{align}}

\def\tr{\mbox{tr }}
\def\beqa{\begin{eqnarray}}
\def\enqa{\end{eqnarray}}

\newcommand{\half}{\textstyle{ \frac{1}{2}}} 
\hypersetup{
    colorlinks=true,       % false: boxed links; true: colored links
    linkcolor=red,          % color of internal links (change box color with linkbordercolor)
    citecolor=blue,        % color of links to bibliography
    filecolor=magenta,      % color of file links
    urlcolor=blue           % color of external links
}

\begin{document}

\title{Hidden topological angles  and  Lefschetz thimbles }
\author{Alireza Behtash}
\email{abehtas@ncsu.edu} 
\author{Tin Sulejmanpasic}
\email{tsulejm@ncsu.edu}
\author{Thomas Sch\"{a}fer}
\email{tmschaef@ncsu.edu}
\author{Mithat \"{U}nsal}
\email{unsal.mithat@gmail.com} 
\affiliation{Department of Physics, North Carolina State University, 
Raleigh, NC 27695, USA}

\begin{abstract}

We demonstrate  the existence of  hidden topological angles (HTAs) in a 
large class of quantum field theories and quantum mechanical systems. 
HTAs are distinct from theta-parameters in the lagrangian. They arise as 
{\it invariant angle} associated with saddle points of the complexified
path integral and their descent manifolds (Lefschetz thimbles). 
Physical effects of HTAs become most transparent upon analytic continuation in 
$n_f$ to non-integer number of flavors, reducing 
in the  integer $n_f$ limit   to a $\mathbb Z_2$ valued phase difference 
between dominant saddles.
In ${\cal N}=1$ 
super Yang-Mills theory we demonstrate the microscopic mechanism for the  
vanishing of the gluon condensate. The same effect leads to an anomalously
small condensate in a QCD-like $SU(N)$ gauge theory with fermions in the 
two-index representation. The basic phenomenon is that, contrary to 
folklore, the gluon condensate can receive both positive and {\it negative} 
contributions in a semi-classical expansion.  In quantum mechanics, a HTA 
leads to a difference in semi-classical expansion of integer and half-integer 
spin particles.

\end{abstract}

\maketitle
{\noindent\bf  Introduction.} 
 Providing a non-perturbative continuum definition of the path integral 
in quantum field theory is a challenging but important problem 
\cite{'tHooft:1977am}.  There is growing evidence that, if an ordinary
integral or a path integral admits a Lefschetz-thimble decomposition  
\cite{Witten:2010zr,Witten:2010cx}
or   resurgent transseries expansion \cite{Dunne:2012zk, Argyres:2012ka, Dunne:2012ae,Dunne:2013ada, Cherman:2014ofa, Dunne:2014bca, Aniceto:2013fka, Misumi:2014jua, Misumi:2014bsa}
then either of these  methods gives 
this long-sought non-perturbative definition. 
If this is indeed the case, then we expect that these new methods will
provide new and deep insight into  quantum field theory and quantum 
mechanics formulated in terms of path integrals. In this article we
introduce a new phenomenon of this kind, the appearance of hidden
topological angles (HTAs). 

 The main prescription associated with the Lefschetz-thimble decomposition 
or the resurgent expansion is the following: Even if an ordinary integral 
or a path integral is formulated over real fields, the natural space that 
the critical points (saddles) $\rho_\sigma$ live in is the complexification 
of the original space of fields. However, the dimension of the critical 
point cycles ${\cal J}_\sigma$ is that of the original space, or half that 
of the complexified field space. For example, for an ordinary integral over 
$N$-dimensional real space, this procedure is $\mathbb R^N 
\longrightarrow \mathbb C^N \longrightarrow  \Sigma^N$, where  $\Sigma^N 
= \sum_\sigma n_\sigma {\cal J}_\sigma$ and $\dim_{ \mathbb R} ({\cal J}_\sigma)
= N$. For $N=1$, this is the well-known steepest descent (stationary phase) 
approximation.  
   
To each  critical point  $\rho_\sigma$ of the complexified action one 
attributes an action, with real and imaginary parts, and with ``weight" 
$e^{-S_\sigma}$. The  imaginary part of the action, ${\rm Im} S_\sigma$  
is an  {\it  invariant angle } associated with the critical point 
$\rho_\sigma$  and its descent manifold   ${\cal J}_\sigma$. If there 
are critical points with the identical real part of the action ${\rm Re} 
S_\sigma$, but different imaginary parts  ${\rm Im} S_\sigma$, then there 
may be subtle effects. 
Indeed, Witten 
recently studied in \cite{Witten:2010cx} the analytic continuation of 
Chern-Simons theory  to non-integer values of the coupling $k$, finding 
subtle cancellations among dominant saddle field configurations in the 
integer $k$ limit, so that the {\it sub-dominant} saddle gives the main
physical contribution. In this work, we show that the effect observed 
by Witten is not an exotic phenomenon, but that it is possibly quite 
ubiquitous, and that it is responsible for a variety of interesting 
physical effects in quantum field theories and quantum mechanics, in which 
the analytic continuation is now to non-integer  ``coupling'' $n_f$, 
which is   the    number of fermionic flavors for integer values.
We 
also show that  the effect is more non-trivial  than a simple cancellation 
between dominant saddles. Indeed, the effect depends on the observable, 
and the dominant saddles may cancels in certain observables, but 
contribute to others.

In a field theory with a topological $\Theta$-angle in the Lagrangian, subtle 
effects may arise at certain values of the  $\Theta$-angle  
\cite{Witten:1997ep,  Unsal:2012zj,Bhoonah:2014gpa,Anber:2013sga}.
For example, at $\Theta=\pi$ in $SU(2)$ gauge theory there is a cancellation 
of leading order saddle contribution to the mass gap \cite{Unsal:2012zj}.  
In this work we study a  more exotic phenomenon, which is due to a hidden 
topological angle not  explicitly present in the lagrangian. We define a
HTA as  the phase associated with a saddle point in the complexified field 
space. Below, we will provide examples of this phenomenon in  
${\cal N}=1$ super Yang-Mills theory (SYM),  
certain QCD-like theories, and the quantum mechanics of a particle with spin. 
We also note that a HTA is different from the discrete theta angles
discussed recently  \cite{Gaiotto:2014kfa}, which comes about as one 
changes the global gauge group. 
In contrast, HTAs are present for any gauge group.    
In the examples discussed below we find that for integer values of the 
number of fermions $n_f$ there is a $\mathbb Z_2$  hidden topological 
structure.

\begin{figure}[ht]
\begin{center}
\includegraphics[angle=0, width=.48\textwidth]{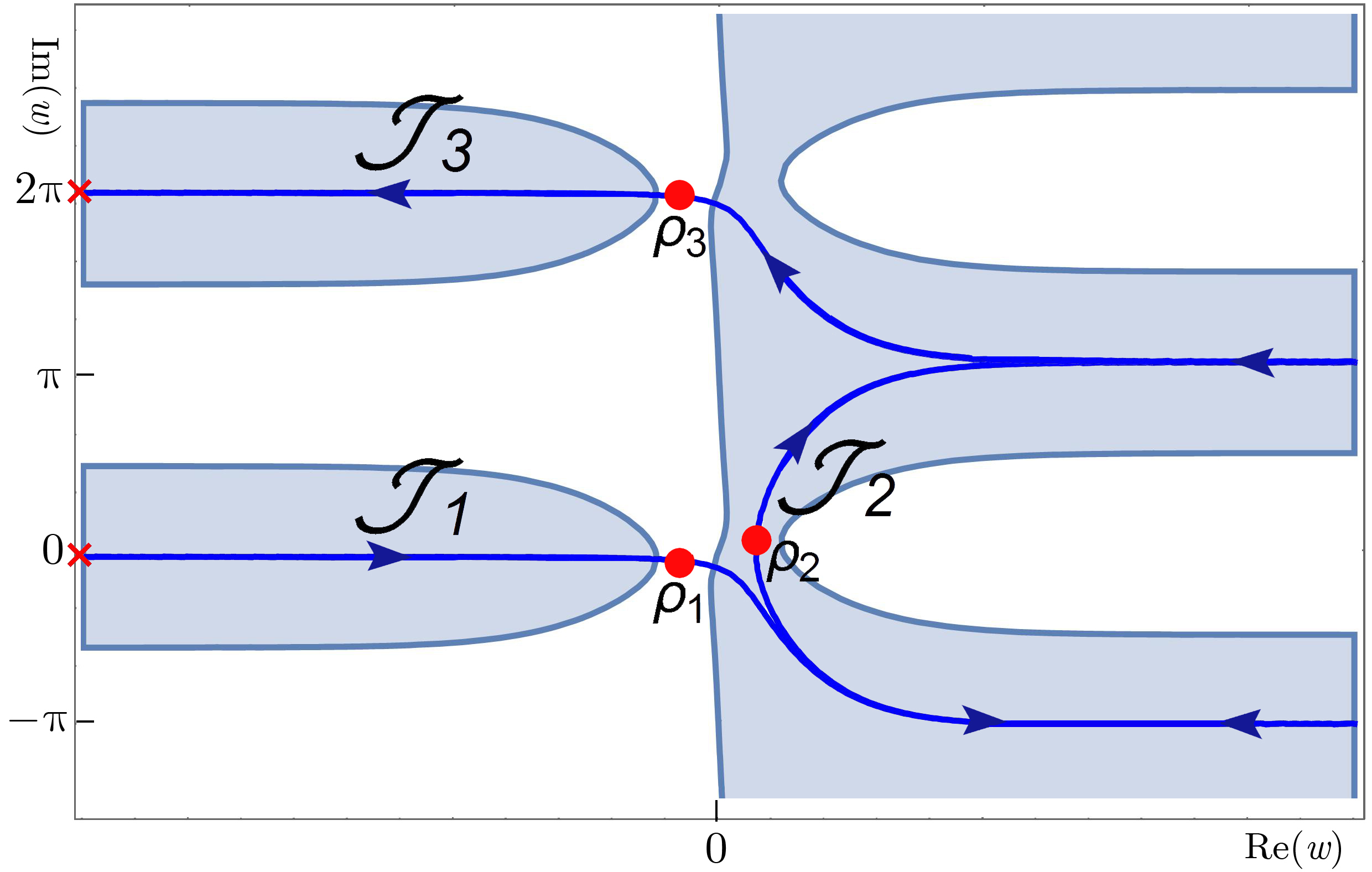}
\caption{The blue areas show ``good regions" in which the integrand 
falls sufficiently rapidly at infinity to guarantee convergence.
The red dots give the locations of the saddle points, and the blue 
contours are the Lefschetz thimbles. If the  boundary of integration 
is $(- \infty, -\infty + 2\pi i)$, then the  Lefschetz decomposition is 
$\mathcal{J}_1+\mathcal{J}_2+\mathcal{J}_3$. Here, $\rho_1$ and $\rho_3$ 
are equally dominant saddles over $\rho_2$, but there is an over-all 
phase difference between  the dominant saddles leading to a subtle 
cancellation for integer $k$.}
\label{Thimbles}
\end{center}
\end{figure}

{\vspace{0.5cm}\noindent\bf A  prototype  in ordinary integration.} 
An elementary example that provides some intuition for field theory is the 
following. Consider the analytic continuation of the Bessel function 
to non-integer order, and describe the contour that appears in the 
integral representation in terms of  Lefschetz thimbles. In one complex 
dimension the Lefschetz thimble is defined as a stationary phase manifold:
${\rm Im} [S(w) - S(w_n)]=0$  where $w_n$ is a critical point on the 
contour. The integral is 
$I(k,\lambda)=\int_{C_w}dw e^{2\lambda\sinh(w)+kw} $
for complex $k$, $\lambda$. In a certain regime of the analytic continuation, 
discussed in \cite{Witten:2010cx}, the integral can be expressed in terms of 
three cycles, ${\cal J}_i, i=1, 2,3$ associated with saddles $\rho_{i}$, so 
that ${C_w} = {\cal J}_1 + {\cal J}_2 +{\cal J}_3$, see Fig.~\ref{Thimbles}.
The  sum  of the three thimble contribution  gives 
\begin{align} 
 (1+  e^{ 2\pi i (k + \frac{1}{2}) }) e^{-S_1}  + e^{-S_2}
 \label{toy-cancel}
\end{align}
where $|e^{-S_1}| =|e^{-S_3} | \gg e^{-S_2}$, i.e, $\rho_1$ and $\rho_3$ 
are dominant over $\rho_2$. However, they have a relative phase, and the 
contribution of these two dominant saddles cancel each other exactly for  
integer $k$. The mechanism described above is  an intuitive  example of 
a mechanism operative in Chern-Simons theory by using analytic continuation, 
providing confidence for the utility of the idea of analytic continuation 
of path integrals. Other examples are discussed in
\cite{Kanazawa:2014qma,Tanizaki:2014tua,Tanizaki:2014xba}.
We will perform a similar analytic continuation in 
$n_f$, the number of fermion flavors in the theory. 

{\vspace{.25cm}\noindent\bf Picard-Lefschetz equation and invariant angles.}
The  definition of the Lefschetz thimble based on stationary phase, ${\rm Im} 
[S(w) -S(w_n)]=0$, is only satisfactory for a one-dimensional integral (it 
provides one real condition on a one-complex dimensional space). In $n$
complex dimensions, where $n=\infty$ corresponds to field theory or quantum 
mechanics, this condition defines a co-dimension one (real dimension $2n-1$ 
space), which is not the desired $n$ real dimensional space. Instead, one 
needs $n$ real conditions to define the thimble. Guided by these observations, 
Witten used complex gradient flow equations, the Picard-Lefschetz equations,
to describe the Lefschetz-thimbles. In a theory with a field $\varphi$ and 
action $S(\varphi)$, this amounts to 
\begin{align}
\frac{\partial \varphi }{\partial \tau} = 
      -  \frac{ \delta \bar S }{ \delta \bar \varphi}\, , \qquad 
\frac{\partial  \bar \varphi }{\partial \tau} = 
      -  \frac{ \delta  S }{ \delta \varphi}  \, ,
\label{PLW}
\end{align}
where $\tau$ is the flow time. Using  \eqref{PLW} and the chain rule, 
\begin{align}
\frac{\partial  {\rm Im} [S] }{\partial \tau} =   
\frac{1}{2i} \left( \frac{\delta S } {\delta \varphi} 
     \frac{\partial \varphi }{\partial \tau} 
 -   \frac{ \delta \bar S }{ \delta \bar \varphi}   
     \frac{\partial\bar \varphi }{\partial \tau}    \right) = 0\, ,
\label{PLW-2}
\end{align}
meaning that  ${\rm Im} [S(\phi)] ={\rm Im} [S(\phi_n)]$ is {\it invariant} 
under the flow.    
In a quantum field theory (QFT), or in quantum mechanics (QM), in which  
semi-classical saddle proliferates (an example is the instanton gas),  
${\rm Im} [S(\phi_n)] $ will appear as a genuinely new phase in  the
effective field theory. This is the HTA phenomenon. 

The  integration in the complexified field space is infinite dimensional. In the background of non-perturbative saddles, this space usually factorizes into  finite dimensional zero and  quasi-zero modes directions  and  infinite dimensional gaussian modes.  The HTA can be calculated  by an exact integration over the  complexified finite dimensional quasi-zero mode directions in the  field space, dictated by the finite dimensional version of the Picard-Lefschetz theory.  

{\vspace{.5cm}\noindent\bf Hidden topological angle in 4d ${\cal N}=1$ SYM:} 
Consider ${\cal N}=1$ SYM on $\mathbb R^3 \times S^1_L$, where $S^1_L$ is 
a circle with period $L$. We use supersymmetry preserving boundary conditions
and take the small $L$ limit in order to be able to use semi-classical
methods.  According to the trace anomaly relation the gluon condensate 
$\langle {\textstyle \frac{1}{N}} {\rm tr} F^2_{\mu \nu} \rangle $ determines
the vacuum energy: ${\cal E}_{\rm vac} = \langle \Omega | T_{00} |\Omega\rangle 
= \frac{1}{4} \langle \Omega |  T_{\mu \mu}   | \Omega \rangle   
= \frac{1}{4} \frac{\beta(g)}{g^3} \langle {\rm tr}\, F^2_{\mu \nu}\rangle$.
This implies that the gluon condensate can serve as an order parameter for 
supersymmetry breaking.  The vacuum energy density, and hence the condensate, 
vanishes to all orders in perturbation theory in supersymmetric theories.
Since supersymmetry is known to be unbroken, the gluon condensate must be 
zero non-perturbatively as well. In the semi-classical limit this result 
appears mysterious, because all contributions appear to be positive. The 
reason is that in euclidean space the fermion determinant is positive 
definite, and $ {\rm tr} F^2_{\mu \nu}$ is also positive definite. This 
implies that the gluon condensate is the average of a positive observable 
with respect to a positive measure \cite{Shifman:1978bx}. Then, how does 
the vanishing of the $ \langle {\rm tr}\, F^2_{\mu \nu} \rangle$ take place 
from a semi-classical point of view?

\begin{figure}[ht]
\begin{center}
\includegraphics[angle=0, width=.52\textwidth]{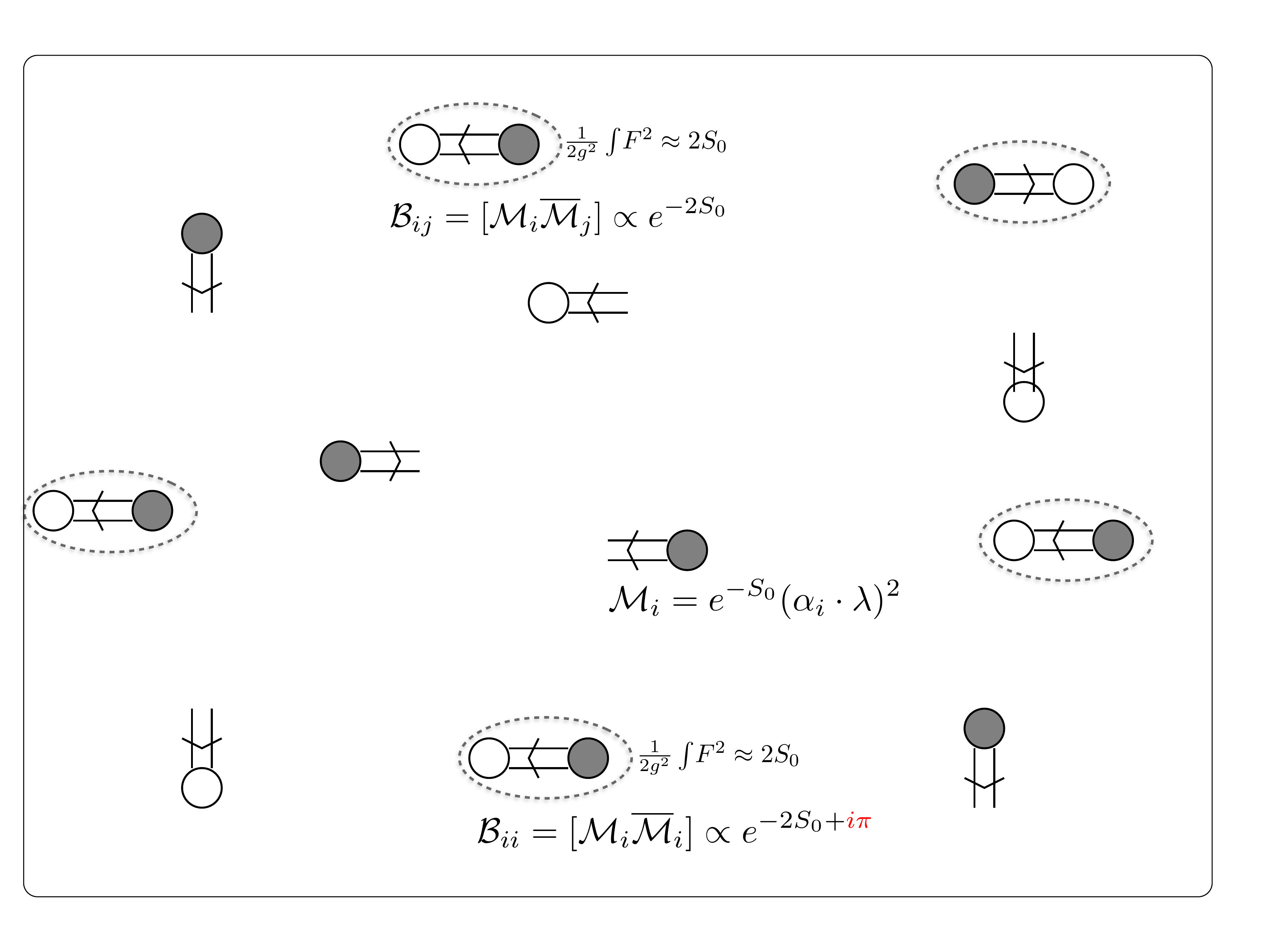}
\caption{A snap-shot of the euclidean vacuum of ${\cal N}=1$ SYM 
on small $\mathbb R^3 \times S^1_L$. Both neutral and magnetic bions 
carry action $2S_0$, but their contribution to gluon condensate cancels 
exactly because of the presence of a HTA, a $\pi$-phase difference between 
the two saddles.}
\label {Thimbles-SYM}
\end{center}
\end{figure}

  We  address this question in the regime of small, but {\it finite} 
radii on  $\mathbb R^3 \times S^1_L$.   To do so, recall the Euclidean 
realization of the 
vacuum of the theory on small $\mathbb R^3 \times S^1_L$, depicted in 
Fig.~\ref{Thimbles-SYM} for the center-symmetric point of the Wilson 
line on the Coulomb branch.  The vacuum is, primarily, a dilute gas  
of semi-classical one- and two-events: monopole-instantons 
\cite{Lee:1997vp,Lee:1998bb,Kraan:1998pm,Davies:2000nw} and bions 
\cite{Unsal:2007jx,Poppitz:2011wy, Argyres:2012ka, Anber:2011de}.   
These are: 
\begin{itemize}
\item[a)] monopole-instantons, ${\cal M}_i =  e^{-S_0} 
           (\alpha_i \cdot \lambda)^2 $, 
\item[b)]   magnetic bions, ${\cal B}_{ij}= [{\cal M}_i \overline {\cal M}_j]
      = e^{-2S_0}\ldots  $,       %$ \forall \hat A_{ij} <0$.   
\item[c)]   neutral bions,  ${\cal B}_{ii}= [{\cal M}_i \overline {\cal M}_i]
      = e^{-2S_0 +i \pi}\ldots$.   %$ \forall \hat A_{ii} >0$.  
\end{itemize}
where $\alpha_i,\;i=1, \ldots, N$ are simple roots complemented with 
the affine root $\alpha_N$, and ${\cal B}_{ij}$ and ${\cal B}_{ii}$ are 
non-vanishing $\forall \hat A_{ij} <0$, and  $\forall \hat A_{ii} >0$ 
entries of the extended Cartan matrix, respectively. The monopole action 
is $S_0= \frac{8 \pi^2}{g^2N}$. For small $L$ the coupling is small, the 
action is large, and fluctuations are suppressed. The $2N$ fermion zero 
modes of the 4d instanton are distributed uniformly as $(2,2,\ldots, 2)$ 
to  monopoles  ${\cal M}_i$. 
	
At leading order $O(e^{-S_0})$ in the semi-classical limit, each 
monopole-instanton has two fermion zero modes and therefore they do 
not contribute to the gluon condensate.  Two-defects {\it do contribute} 
to the gluon condensate.  For the sake of making the analogy with the 
toy example \eqref{toy-cancel} explicit, let us   consider analytic 
continuation  away from  $n_f=1$. The density of both types of 2-defects 
is the same, of order $O(e^{-2S_0})$. However, there  is an extra $(4n_f-3) 
\pi$ phase (invariant angle) associated with the neutral bion saddle/thimble:
\begin{align} 
{\rm Arg} ( \mathcal{J}_{ {\cal B}_{ii}} )  
    = {\rm Arg} (\mathcal{J}_{ {\cal B}_{ij}} )  +  (4n_f-3) \pi\, . 
\end{align}
Consequently, in contrast to the folklore regarding the positivity
of the gluon condensate, the contributions of the two types of 2-defects 
to the gluon condensate cancel:  
\begin{align}
\label{cancel}
L^4  \langle {\textstyle \frac{1}{N}} {\rm tr} F^2_{\mu \nu} \rangle  
   &=  0  \times n_{ {\cal M}_{i}} +   
    (n_{ {\cal B}_{ij}} +   e^{i  (4 n_f -3) \pi}n_{ {\cal B}_{ii}})  = 0 \, . 
\end{align}
at a physical integer value of the parameter, $n_f=1$, similar to 
\eqref{toy-cancel}. This is the microscopic mechanism for the vanishing of 
the gluon condensate as well as the vacuum energy in ${\cal N}=1$ SYM.  
The two contributing bion-thimbles are charged oppositely under the 
${\mathbb Z}_2^{\rm HTS}$,  and cancel each other out.

 The  difference with respect to the toy integral and the cancellation 
in analytically continued Chern-Simons theory is the fact that this 
cancellation is observable dependent. In fact, the combination of the 
neutral and magnetic bions, despite giving vanishing contribution to 
gluon condensate, is responsible for the formation of a mass gap.  To 
see this consider the effective lagrangian for the low energy bosonic
modes. As an example, we will use $SU(2)$ gauge theory. Let $\phi$ 
denote the fluctuation of the Wilson line  around the center symmetric 
minimum and $\sigma$ denote the dual photon. The  bosonic potential 
induced by 2-defects is \cite{Poppitz:2012sw}
\begin{align}
V(\sigma, \phi) &=   - \left( {\cal B}_{12} +  {\cal B}_{21} 
   +  {\cal B}_{11} +  {\cal B}_{22} \right)  \cr
& \sim  e^{-2S_0}  \left( -\cos {2\sigma}  -  e^{i\pi}  \cosh{2\phi} \right)\, .
\end{align}
We observe that the factor $e^{i \pi }$ responsible for the vanishing  
$ \langle {\textstyle \frac{1}{N}} {\rm tr} F^2_{\mu \nu} \rangle$ is 
also responsible for the (positive and un-suppressed) mass gap of the 
$\phi$-fluctuations and stabilizes center-symmetry. The HTA explains
both the vanishing of the gluon condensate and the non-tachyonic 
nature of fluctuations of the Polyakov line. It is also not particular to supersymmetric theory, 
as we discuss next.

{\vspace{0.5cm}\noindent\bf  QCD(AS/S):} 
In a typical confining asymptotically free $SU(N)$ gauge theory, the 
``natural"  scaling of the  (properly normalized)  gluon condensate is 
$O(N^0)$: $\big\langle\textstyle\frac{1}{N}{\rm tr} F_{\mu \nu}^2\big\rangle\propto 
N^0\Lambda^4$. It is  natural to expect that  the vanishing of the gluon 
condensate is special to the supersymmetric theory. This is not the case. 
There exists an {\it exact} large-$N$ orientifold/orbifold  equivalence 
between ${\cal N}=1$ SYM and QCD(AS/S) \cite{Armoni:2003gp}, proven in  
\cite{Kovtun:2005kh}. Here, AS/S refers to fermions in 
anti-symmetric/symmetric two-index representations. The large-$N$  
equivalence implies that the gluon condensate in QCD(AS/S) is zero at 
leading order in the $1/N$ expansion, and must scale as \cite{Unsal:2007fb}:
\begin{align}
 \big \langle  \textstyle \frac{1}{N} \tr F_{\mu \nu}^2  \big \rangle^{\rm QCD(AS/S) }
   = \frac{1}{N} \Lambda^4\, , 
   \label{suppressed}
\end{align}   
in sharp contrast with the ``natural" value. This result is counter-intuitive, 
but it is a rigorous consequence of the large $N$ equivalence. However, as 
in the supersymmetric case, there is no known semi-classical explanation.

 Again, we can understand the result based on the presence of HTAs. To 
achieve this, we use the framework of deformed Yang-Mills theory, and 
add AS representation fermions (a similar analysis holds for QCD(S)). 
In QCD(AS), the Atiyah-Singer index theorem implies that the number of 
fermion zero modes of a 4d instanton is $2N-4$. There are $N$ types of
monopole-instantons, with the number of fermion zero modes distributed 
as $(2,2,\ldots,2,0,0)$ in a center-symmetric background. The difference 
with respect to ${\cal N}=1$  SYM is that 2 out of $N$ monopole-instantons 
do not possess fermi zero modes. Therefore, at leading order, $O(e^{-S_0})$,
in the semiclassical expansion, $N-2$ monopoles do not contribute to the 
gluon condensate and {\it only} two do, giving a positive contribution
proportional to $1/N$. At second order in the semiclassical expansion,
$O(e^{-2S_0})$, there are magnetic and neutral bions that can contribute 
to gluon condensate. Their contribution cancels at leading order in $N$, 
analogous to  SYM, leading to  \eqref{suppressed}.

 In the older literature on QCD \cite{Callan:1977gz,Shifman:1978bx,Schafer:1996wv}, 
it was assumed that in the semi-classical limit gluon condensate is 
proportional to the instantons density. This was based on the rationale 
that a single instanton contributes  a finite and positive amount, 
$\frac{1}{2g^2} \int \tr F_{\mu \nu}^2=\frac{8\pi^2}{g^2}$, and that the condensate 
can be attributed to 4d instantons with a positive weight, $\big\langle
\textstyle \frac{1}{N} \tr F_{\mu \nu}^2 \big \rangle \propto n_I$. In the calculable 
small $S^1_L$ regime, we see that this is incorrect in at least two ways: 
{\it i)} Instantons are sub-leading, i.e. $O(e^{-N})$, {\it ii)} The weight 
of the saddles can be both positive (decreasing energy) and negative (increasing energy). Our work is the first 
example in which a contribution to condensate has a negative component.

{\vspace{.5cm}\noindent\bf Quantum mechanics:}  
In order to show the generality of hidden topological angles, we also  
consider the quantum mechanics of a particle with position $x(t)$ and 
internal spin $(\half)^{n_f}$. The euclidean Lagrangian is that of a bosonic 
field $x(t)$ coupled to $n_f$ fermionic fields $\psi_i$: 
\begin{equation}
{\cal L}_E = \left( \half\dot x^2 + \half (W')^2 
            +  i (\bar\psi_i \dot\psi_i  + W'' \bar \psi_i\psi_i )  \right)
\label{lag}
\end{equation}
For $n_f=1$, this theory is supersymmetric \cite{Witten:1981nf}. If we choose $W(x)$ 
to be a periodic function, for example $W(x)= \cos x$, we may identify 
$x= x+ 2\pi$ as the same physical point, corresponding to a 
 particle on a circle, rather than in an infinite lattice. The system 
contains two types of instantons, 
\begin{align}
{\cal I}_1: [0 \rightarrow \pi], \qquad   
{\cal I}_2: [0 \rightarrow -\pi]  \, . 
\end{align}
Here,  $ {\cal I}_2$  is an instanton (not an anti-instanton), because 
it satisfies the same BPS equation (or gradient flow equation, if $W$ 
is viewed as a Morse function) as ${\cal I}_1$. 
 
 Because of spin, instantons do not contribute to the vacuum energy. A
non-vanishing contribution arises from correlated two-events.  This
parallels the 4d field theory on ${\mathbb R}^3 \times S^1$, and provides 
a simple system in which the effect of the HTA is not contaminated by first 
order instanton effects. Following \cite{Balitsky:1985in,Poppitz:2012nz},
we find that the amplitudes of the two-events are given by  
$ [{\cal I}_{1}  \overline {\cal I}_{1} ]   
= [{\cal I}_{2}  \overline {\cal I}_{2} ]  \propto e^{ i \pi n_f} e^{-2S_{I}}$, 
and  
$ [{\cal I}_{1}  \overline {\cal I}_{2} ]  
= [{\cal I}_{2}  \overline {\cal I}_{1} ] \propto e^{-2S_{I}} $.  
Due to the difference in the invariant angles between the two-saddles/thimbles
we find
\begin{align}
 \label{thimble-2}
  {\rm Arg} (\mathcal{J}_{[{\cal I}_{1}  \overline {\cal I}_{1} ]}  )  
= {\rm Arg} (\mathcal{J}_{[{\cal I}_{1}  \overline {\cal I}_{2} ]}  )  + n_f \pi \, . 
\end{align}
The non-perturbative contribution to the ground state energy takes the form: 
 \begin{align}
& \Delta E_0^{\rm np} = ( -2 -2  e^{ i \pi  n_f}  ) e^{-2S_{I}} 
\end{align} 
While the $[{\cal I}_{1}\overline {\cal I}_{2} ]$ molecule behaves in the 
expected manner and decreases the ground state energy, the $[{\cal I}_{1}  
\overline {\cal I}_{1}]$ molecule is sensitive to the HTA governed by the 
spin and increases the ground state energy for odd $n_f$  (half-integer 
spin) while decreasing it for even-$n_f$ (integer spin). 
In the case
 $n_f=1$, \eqref{thimble-2}  is the  microscopic reason for the 
non-perturbative vanishing of the  ground state energy. We note that 
this system, despite having Witten index zero  \cite{Witten:1982df}, $I_W={\rm tr}[(-1)^F]=0$,  
has unbroken supersymmetry, and two supersymmetric ground states. 
  
  There are two additional interesting features of this system. The first 
is related to the fact that one can introduce a topological angle into the 
Lagrangian, $i \frac{\Theta}{2 \pi} \int\dot{x}\, d\tau$, and unlike the case of supersymmetric 
gauge theory, the $\Theta$-angle is physical, and alters the spectrum of the 
theory. Since $I_W=0$, the supersymmetry of this system is fragile. The vacuum 
energy can be written as    
\begin{align}
& \Delta E_0^{\rm np} =( -2 \cos \Theta-2  e^{i \pi n_f }  ) e^{-2S_{I}} 
 \label{thimble-3}
\end{align} 
which, for the supersymmetric theory $(n_f=1)$, takes the form: 
\begin{align}
& \Delta E_0^{\rm np} =( -2 \cos \Theta +2  ) e^{-2S_{I}}  
   = 4 \sin^2 \textstyle{\frac{\Theta}{2}} e^{-2S_{I}}   \geq 0
 \end{align} 
meaning supersymmetry is dynamically broken for $\Theta\neq 0$. Note that 
the energy remains positive semi-definite, which is a consequence of the 
supersymmetry of the Hamiltonian. The physical reason for $E>0$ in the 
case $\Theta\neq 0$ is that the  $\Theta$-angle is equivalent to feeding 
momentum into the system.  Because of supersymmetry, bosonic/fermionic ground 
state pair is  lifted simultaneously by the insertion of momentum, leading to 
a non-vanishing ground state energy.

 The second unusual feature is that the theory has Witten index $I_W=0$ for 
any value of $\Theta$, but that the reason for $I_W=0$ differs in the two
cases. For $\Theta=0$, we get $I_W=1-1=0$, where the two contributions arise 
from the bosonic/fermionic sectors of the Hilbert space, and supersymmetry 
is unbroken. In the second case, $\Theta\neq 0$, we get $I_W=0-0=0$ and
supersymmetry is broken.

 One may speculate that the invariant angles are related to Berry phases 
\cite{Berry:1984jv}, realized in terms of Euclidean saddles, at least in 
the case of quantum mechanics.
Since $(\half)^{n_f} = \bigoplus_{S}{\rm mult}(S) S$, where ${\rm mult}(S)$ is 
the multiplicity, we can rewrite the path integral over the Grassmann variables
as spin path integrals \cite{Stone:1988fu} $Z= \sum_{S}{\rm mult}(S) Z^{(S)}$ 
where 
\begin{align}
\label{eq:spf}
 {Z}^{(S)}  &= \int Dx D(\cos\theta) D\phi\; 
    e^{ -{\cal L}_E   + i  \frac{ \Theta}{2 \pi}  \int \dot x d \tau  }  \\
{\cal L}_E &=
  \textstyle{ \frac{1}{2}} (  \dot{x}^2 + (W')^2) +  S W''\cos \theta   
        +  iS (1- \cos\theta ) \dot{\phi}   \nonumber
\end{align}
where  $(\theta, \phi) \in {\bf S}^2$ parameterize the Bloch sphere. There
are two spin dependent interactions, a ``magnetic field" $W''$-spin coupling, 
and the Wess-Zumino term, or Berry phase action. In this language
\eqref{thimble-2} should be replaced by %$n_f\pi= 2S\pi$. 
${\rm Arg} (\mathcal{J}_{[{\cal I}_{1}  \overline {\cal I}_{1} ]}  )  
={\rm Arg} (\mathcal{J}_{[{\cal I}_{1}  \overline {\cal I}_{2} ]}  ) + 2S\pi$.
This  distinguishes  half-integer and integer spin particles, similar to  
anti-ferromagnets in one-spatial dimension  \cite{Haldane:1983ru}, leading 
to qualitative differences.

{\vspace{.5cm}\noindent\bf Conclusion:}
We have provided several examples of $\mathbb Z_2$ hidden topological
angles, associated with the saddle point manifolds that appear in the 
complexified path integral, and  have shown that these angles lead
to crucial physical effects. 
We anticipate that HTAs 
will have crucial impact on the semi-classical analysis of many 
interesting quantum field theories and quantum mechanical systems. 
Our examples also show that in an attempt to perform lattice simulations 
using Lefschetz thimbles, e.g.,  
\cite{Cristoforetti:2012su,Cristoforetti:2013wha,Fujii:2013sra,Aarts:2014nxa},  
all thimbles whose multipliers are non-zero must be carefully summed over   
to correctly capture the dynamics of the theory.

{\it Acknowledgments.}
%%%%%%%%%%%%%%%%%%%%
We thank G. Dunne, G. 't Hooft, P. Argyres, and E. Witten  for useful 
comments and discussions. We acknowledge support from DOE grants 
DE-FG02-03ER41260 and DE-SC0013036.

\bibliographystyle{apsrev4-1}
\bibliography{HTAbibliography}

\end{document}